\documentclass[prb,twocolumn,floatfix]{revtex4-2}
\usepackage{epsfig}
\usepackage{graphicx}
\usepackage{amsfonts,amssymb}
\usepackage{amsmath}
\usepackage{color}
\usepackage{times}
\usepackage{flushend}
\usepackage{xcolor}
\usepackage{hyperref}

\hypersetup{
    colorlinks=true, % Enable colored links
    linkcolor=blue, % Color of internal links (sections, figures, etc.)
    citecolor=blue, % Color of citation links (bibliography)
    filecolor=magenta, % Color of file links
    urlcolor=blue, % Color of URL links
}

\begin{document}

\title{Dynamics in the nonequilibrium energy landscape of a frustrated
Mott insulator}

\author{Sankha Subhra Bakshi, Tanmoy Mondal and Pinaki Majumdar}

\affiliation{Harish-Chandra Research Institute
 (A CI of Homi Bhabha National Institute), 
Chhatnag Road, Jhusi, Allahabad 211019
}
\date{\today}

\begin{abstract}
In a Mott insulator, a laser pulse with frequency tuned to the gap scale can 
create a holon-doublon plasma, suppressing the magnetic moment ${\vec m}_i$ 
and destroying magnetic order. While this disruptive effect is well established 
experimentally on a square lattice, we investigate the effect of laser pumping 
on the triangular lattice, where geometric frustration leads to a richer set of 
ordering possibilities. We work with the Mott-Hubbard problem at a coupling 
where  $120^{\circ}$ order is just stable and employ spatio-temporal mean field 
dynamics to study the pump response. Moderate pump amplitude just leads to 
reduction of $120^{\circ}$ order, but at larger amplitude the suppression of  
$120^{\circ}$ order is followed by the appearance of `spiral order'. On the 
electronic side the density of `excited carriers' $n_{exc}$ in the upper 
Hubbard band increases monotonically with pump amplitude. We show that the long 
time ordering possibilities in the pumped system, e.g, the emergence of spiral 
order, can be inferred from a nonequilibrium `energy landscape'.
We analyse the growth of spiral order by 
using an exact diagonalisation based Langevin equation on large lattices and
discover that the new order can take $\sim 10^3-10^4$ times the electronic  
timescale to appear.  The threefold combination, of mean field dynamics, 
landscape construction, and Langevin dynamics, readily generalises to the 
search for pump induced `hidden order' in other gapped systems. 
\end{abstract}

\maketitle

\section{Introduction:}

The suppression of long range order in correlated systems 
by using a short 
intense laser pulse is well-known~\cite{OS1,OS2,OS3,OS4,OS5}. 
Recently, it has 
been shown that such `pump' pulses can not only suppress an 
existing order but also promote new kinds of order, e.g., 
photoinduced charge density wave~\cite{PICO1,PICO2,PICO3,PICO4,
PICO5,PICO6}, 
orbital order~\cite{PIOO1,PIOO2},
superconductivity~\cite{PISC1,PISC2}, etc. It has also been 
shown that with laser pumping one can temporarily sustain long 
range order beyond the equilibrium critical 
temperature~\cite{PISC1,PISC2,PIHTF}. 
In general, laser pumping offers a new way to 
manipulate quantum matter and Mott insulators are promising 
for photoinduced states since they support 
complex magnetic order, and are also `adjacent' 
to novel metals and superconductors~\cite{MI1,MI2,MI3}.

The simplest realisation of Mott physics is in the single
band  Hubbard model at half-filling.  On a square lattice, 
with nearest neighbour hopping $t_{hop}$,
arbitrarily weak repulsion $U$ leads to a spin density 
wave insulator with  
ordering wavevector ${\bf Q} = (\pi/a, \pi/a)$
\cite{sqr1,sqr2},
where $a$ is the lattice parameter.
Increasing $U/t_{hop}$ leads to growing size of the local moment
and the Mott gap, but the order remains at $(\pi/a, \pi/a)$.
There are no `competing phases' at half filling.
The triangular lattice is more exciting. A variety of tools have 
been used to study it, from mean field theory to density matrix
renormalisation group (DMRG)~\cite{tr1,tr2,tr3,tr4,tr5,tr6}, 
but even at half-filling a consensus is yet to reach. 
However a mean field treatments of this problem~\cite{tr7} show 
(i)~there 
is a correlated non magnetic metal upto some $U_{c1}$, then
(ii)~a magnetic metal with varying 
wavevector between  $U_{c1}$
and $U_{c2}$, then (iii)~the 
$120^{\circ}$ ordered insulator beyond $U_{c2}$.
There is phase competition in this model, unlike 
in the square lattice where the $(\pi, \pi)$ state
always wins.

Experimental and theoretical investigations 
~\cite{review1,review2,ppmott1,ppmott2,ppmott3,ppmott4,
ppmott5,ppmott6,ppmott7,ppmott8,ppmott9,ppmott10,ppmott11} 
on photo-excited Mott insulators have revealed various 
nonequilibrium phenomena, including insulator-to-metal 
transitions \cite{IMT1,IMT2}, suppression and revival of 
long-range order (as observed in Sr$_2$IrO$_4$ \cite{sr2iro41,sr2iro42,sr2iro43}), and metastable 
charge-orbital-spin reordering across materials like 
1T-TaS$_2$ \cite{1ttas2}, FeSe \cite{Fese}, 
and WSe$_2$ \cite{wse}.
Additionally, the role of 
non-equilibrium population dynamics \cite{doublon1,doublon2} 
in these processes has been highlighted. These meta-stable 
states, which cannot be accessed through thermal 
excitation alone, have attracted considerable interest. 
Spectral analysis has further distinguished photo-induced 
metallic states from conventional high-temperature 
metallic phases \cite{ex1,ex2}.

Theoretically dealing with pump induced emergence of order 
is challenging due to several factors. First, the temporal 
evolution needs to cover a wide range of timescales, 
from fast electronic scales to the 
slower scales for magnetic moments and then 
collective timescales associated with domain dynamics. 
Second, spatial correlations must be accounted for, and the 
system size should be large not only to afford high enough resolution 
in momentum space but also to see domain competition arising
in such scenarios.
Third, the nonequilibrium population created by the exciting 
pulse must be taken care of. 
Numerically exact techniques face hurdles here: exact 
diagonalization (ED) \cite{ED1,ED2} is very size-limited, 
dynamical mean field theory (DMFT) 
\cite{DMFT0,DMFT1,DMFT2,DMFT3,DMFT4,DMFT5,DMFT6}
misses the spatial 
correlations, and DMRG \cite{DMRG1, DMRG2} is limited to mainly one 
dimension. 
Phenomenological Ginzburg-Landau \cite{GL1,GL2,GL3} approaches lack 
a microscopic description and do not handle the
excited electronic degrees of freedom faithfully.

In this paper, we attempt to produce a comprehensive 
understanding of a pump induced `suppression-emergence'
problem by studying the triangular lattice Mott insulator
in the regime of $120^{\circ}$ order.
We use a combination of methods.
Our primary microscopic tool is mean field dynamics (MFD),
which arises from factorising the 
Heisenberg equation of motion
for the density operator
 $\hat{\rho}_{ij}^{\sigma\sigma'}(t)= c_{i\sigma}^{\dagger}(t) 
c_{j\sigma'}(t)$.
It allows us to track the spatio-temporal evolution of the magnetic
moment, ${\vec m}_i(t)$, and also infer electronic properties.
The expectation is that $120^{\circ}$ order will be suppressed
and ultimately destroyed as the pump strength $E_0$ is increased.
While this does happen, we find that complete suppression
of  $120^{\circ}$ order is followed by the emergence of
a new kind of order! 
On the electronic side, there is 
a growing upper Hubbard band (UHB) population, i.e,
double occupancy, stable at long times, 
as the pump strength is increased.
This density of `excited electrons' $n_{exc}$ turns out
to be a key player in the system. Extracting $n_{exc}$ as
a function of pump strength from MFD we use it as 
input to a variational calculation (VC) to map out a
nonequilibrium `energy landscape' that explains the 
order seen within MFD. 
Finally, to study the domain dynamics involved in the
emergence of spiral order
we construct a Langevin dynamics
(LD) scheme that incorporates the excited 
electron population $n_{exc}$.
Setting $\tau_0 = 1/t_{hop}$ as the
reference timescale in the problem,
our main results are the following:

(i)~{\it Appearance of new order:} 
A weak pulse just reduces the magnitude of $120^{\circ}$ order 
but for $E_0 > E_0^{c1}$ the suppression of this order 
is followed by the emergence of competing domains of spiral order. 
For $E_0 > E_0^{c2} > E_0^{c1}$ the magnetic moment 
itself is quenched, all order is lost, and we see a paramagnet. 

(ii)~{\it Excited electron population:}
Pump excitation leads to double occupancy and an associated
upper Hubbard band population $n_{exc}$. 
This population stabilises on a timescale of $\sim 50 \tau_0$ 
independent of $E_0$, and remains constant despite significant
changes in the spin configuration, and associated electronic 
density of states, over time.

(iii)~{\it Nonequilibrium phase diagram:}
The association of new order with a high energy population 
$n_{exc}$ is confirmed by a variational calculation. 
We map out a nonequilibrium $U-n_{exc}$ 
phase diagram showing transitions from $120^{\circ}$ order
to spirals and then a local moment paramagnet.

(iv)~{\it Domain dynamics:}
While MFD hints at new order and variational calculations confirm
them, the actual approach to that state involves the growth of 
competing domains. Using Langevin dynamics on large spatial scales we
find that the growth timescale is sensitive to 
pumping strength and can vary between $\sim (10^3-10^5)\tau_0$.

This paper is organized as follows. In Section II we
explain the model, the equations of motion that we
solve and the indicators. 
Section III recapitulates the results at equilibrium.
Section IV presents results on order parameter dynamics
and excited electron population obtained via MFD. 
In Section V we construct a nonequilibrium 
energy landscape incorporating the excited electron 
population, while Section VI uses
Langevin dynamics to capture the growth of spiral
ordered domains and estimate the associated 
timescales. Section VII discusses some
methodological issues. We then conclude.

\section{Model and Method}

We investigate the single-band Hubbard model at half-filling on a 
triangular lattice, described by the Hamiltonian:
\begin{equation}
H = \sum_{\langle ij \rangle, \sigma} t_{ij} c_{i\sigma}^\dagger 
c_{j\sigma} + U \sum_{i} n_{i\uparrow} n_{i\downarrow},
\end{equation}
where \( t_{ij} \) is the nearest-neighbor hopping amplitude 
(\(-t_{hop}\)) and \( U \) is the interaction strength. 
The operators \( c_{i\sigma}^\dagger \) and \( c_{j\sigma} \) 
are fermionic creation and annihilation operators, respectively, 
and \( n_{i\sigma} = c_{i\sigma}^\dagger c_{i\sigma} \) is the 
number operator.
To incorporate a classical light pulse, we apply the Peierls 
substitution to the hopping parameter:
$
t_{ij} \rightarrow t_{ij} \exp\left(i \int_{\vec{R}_i}^{\vec{R}_j} 
\vec{A}(t) \cdot \vec{dr}\right),
$
where \( \vec{A}(t) \) is the vector potential. The electric field 
is given by \( \vec{E} = -\frac{\partial \vec{A}}{\partial t} \), 
and for the light pulse, it is:
\begin{equation}
\vec{E}(t) = \vec{E}_0 \exp\left(-\frac{(t - t_0)^2}{2 \tau_p^2}\right) 
\sin(\omega_p t),
\end{equation}
with \( \vec{E}_0 \) as the amplitude, \( t_0 \) the center time, 
\( \tau_p \) the pulse width, and \( \omega_p \) the frequency.

\subsection{Mean field dynamics}

At half-filling, the interacting problem can be mapped to a non-interacting 
problem coupled to a magnetic background using the Hubbard-Stratonovich 
transformation. The effective Hamiltonian then takes the form~\cite{HSF1,HSF2}:
\begin{equation}
    H_{SF} = \sum_{ij,\sigma} t_{ij} c^{\dagger}_{i\sigma} c_{j\sigma}
    - 2U \sum_{i} \hat{\vec{m}}_i \cdot \hat{\vec{s}}_i 
    + U \sum_{i} \hat{\vec{m}}_i^2,
\end{equation}
where \(\hat{\vec{s}}_i = \frac{1}{2} \sum_{\sigma,\sigma'} 
\vec{\tau}_{\sigma\sigma'} c^{\dagger}_{i\sigma} c_{i\sigma'}\), 
and \(\vec{\tau}\) are the Pauli matrices. The background field 
\(\vec{m}_i(t)\) represents the local magnetization and is determined 
self-consistently by
$ \vec{m}_{i} = \langle \hat{\vec{s}}_i \rangle $.
We can write the Heisenberg equation 
for the density operator \(\rho_{ij}^{\sigma\sigma'}(t) = 
\langle c^{\dagger}_{i\sigma}(t) c_{j\sigma'}(t) \rangle\), 
and `close it' using the condition above \cite{MFD1,MFD2,MFD3,MFD4}:

\begin{equation}
  \begin{split}
   \frac{d}{dt} \rho^{\sigma\sigma'}_{ij} &= i \sum_k \left(t_{ki} 
   \rho^{\sigma\sigma'}_{kj} - \rho^{\sigma\sigma'}_{ik} t_{jk}\right) \\
    &\quad + 2iU \sum_{\gamma} \rho^{\sigma\gamma}_{ij} (\vec{m}_j 
    \cdot \vec{\tau}_{\sigma'\gamma}) - (\vec{\tau}_{\gamma\sigma} 
    \cdot \vec{m}_i) \rho^{\gamma\sigma'}_{ij}
\end{split}
\end{equation}
with \(\vec{m}_{i}(t) = \frac{1}{2} \sum_{\sigma\sigma'} 
\vec{\tau}_{\sigma\sigma'} \rho_{ii}^{\sigma\sigma'}(t)\). 
These are \(4N^2\) first-order differential equations, where 
\(N\) is the number of sites, making the numerical complexity 
scale as \(\mathcal{O}(N^2)\) per time step. This method, known 
as mean-field dynamics (MFD), captures 
the time evolution given an initial state.

We use the 4-th order Runge-Kutta method
to solve this family of equations and set the integration 
timestep $dt=0.01 \tau_0$, where $\tau_0 = 1/t_{hop}$.
We set the lattice distance $a=1$.

\subsection{Nonequilibrium energy landscape}

We find that the pump generates a long time upper Hubbard
band population $n_{exc}(E_0)$ 
defined as the total population occupying 
the states with energy $\omega>0$. 
We construct an electron population function based on the
long time MFD result that reproduces this $n_{exc}$.
This population function is then used as an input 
in constructing an energy function in the space of
ordered configurations, parametrised below:
\begin{equation}
{\vec m}_i = m({\hat x} cos(\vec{q}\cdot{\vec r}_i) +
{\hat y} sin(\vec{q}\cdot{\vec r}_i))
\end{equation}
This configurations enters in
the Hamiltonian $H_{SF}\{{\vec m}_i\}$ which has eigenvalues
$\epsilon_n$ that depend on $(m, q_x, q_y)$. We call this
parameter set $\alpha$. The associated
DOS is $\mathcal{N}(\omega, \alpha) = \sum_n \delta(\omega - \epsilon_n^{\alpha})$.
The energy per site associated with the configuration $\alpha$ is
\begin{equation}
{\cal E}({\alpha}) = \frac{1}{N} \int \omega d \omega ~\mathcal{N}(\omega, \alpha)
P(\omega) + U m^2
\end{equation}
where $ P(\omega) $ is the occupation function extracted
from MFD (see later) and is parametrised uniquely
by $n_{exc}$.
For a given $n_{exc}$, the optimal $m$, $q_x$, $q_y$ can be
obtained by setting ${{\partial {\cal E}}/{\partial m}} = 0,
{{\partial {\cal E}}/{\partial q_x}} = 0 $,
etc, and checking that it indeed is a minimum.
This scheme is a simple generalisation of the equilibrium
scheme, at $T=0$, where one sets $P(\omega) = \theta(- \omega)$. Here, when we talk about the energy 
landscape, we often plot the $\delta \mathcal{E}(\alpha)= \mathcal{E}(\alpha) - \text{min}\left[\mathcal{E}(\alpha)\right]$

\subsection{Langevin dynamics}

While the energy minimsation above gives us a hint about what
ordered states can emerge if a finite $n_{exc}$ is present, it 
does not tell us how that state  dynamically emerges after the
suppression of $120^{\circ}$ order.
To access this we use a Langevin equation directly for the
moments ${\vec m}_i$
\cite{MFD1} instead of considering the complicated
object $\rho_{ij}^{\sigma \sigma'}$. 
Unlike Landau-Lifshitz-Gilbert 
(LLG) evolution, this method allows the magnitude of ${\vec m}_i$ to 
fluctuate. The equation, below, resembles the Langevin equation 
for a classical spin model, with the difference that 
the  `torque' is of electronic origin, and takes into account
the predetermined occupation of excited electronic states.
The Langevin equation, with thermal noise and dissipation satisfying 
the fluctuation-dissipation relation at temperature \(T\) is:
\begin{equation}
    \frac{d\vec{m}_i}{dt} = - \vec{m}_i \times \left\langle 
    \frac{\partial H_{SF}}{\partial \vec{m}_i} \right\rangle
    - \gamma \left\langle \frac{\partial H_{SF}}{\partial \vec{m}_i} 
    \right\rangle 
    + \vec{\xi}_i,
\end{equation}
where \(\gamma\) is the dissipation rate and \(\vec{\xi}_i\) is the 
thermal noise defined as \(\langle \xi_i(t) \xi_j(t') \rangle = 
2\gamma T \delta_{ij} \delta(t-t')\). 
In our case, the initial state in the mean-field dynamics (MFD) 
corresponds to a very low internal temperature 
(e.g., \(T/t_{hop} \sim 10^{-6}\)). However, the pump can 
introduce a doublon-holon background, which acts as a 
thermal bath for the collective modes with an effective temperature
$T$ and dissipation. Given the thermal noise can be treated as
colorless, this dissipation rate 
is selected to comply with the fluctuation-dissipation theorem. 
The microscopic origins of this dissipation—whether from an 
external bath, charge fluctuations, or low-energy phonons—are 
not the focus here and are treated as a phenomenological 
damping parameter.

Numerically, the difficult part is computing 
 \(\left\langle \frac{\partial H_{SF}}
{\partial \vec{m}_i} \right\rangle\) since it requires knowledge
of the system eigenvalues and eigenfunctions in an arbitrary
spin background, i.e., diagonalization of the $2N \times 2N$ system matrix.
This is a ${\cal O}(N^3)$ cost per local update.
We found that the torque on a moment ${\vec m}_i$ 
can be accurately estimated by 
diagonalising a `cluster Hamiltonian' centered on ${\bf R}_i$.
A cluster with size $N_c =19 $, consisting of
the nearest neighbour and next nearest neighbour sites on the
triangular lattice, is adequate to calculate the torque.
This approach reduces the system update cost from $\mathcal{O}(N^3)$ to
$\mathcal{O}(N_c^3N)$. We use the Euler-Maruyama algorithm 
to solve these coupled stochastic differential equations.

\subsection{Parameters}

In this study, we mostly work with $U/t_{hop} = 6.4$ and 
set $t_{hop}= 1$. The initial state is the 120-degree
ordered mean-field ground state.
For the pump response, we use a pulse that has 
a frequency $\omega_p = t_{hop} $
which is close to
the band gap ($\Delta \sim 1.3~t_{hop}$), inducing  
transitions from the lower Hubbard band to the 
upper Hubbard band. The pulse's envelop  contains a few 
oscillations of the electric field ($\tau_p=4~t_{hop}^{-1}$).
We only vary the amplitude 
of the electric field $E_0$ as a control parameter.
The pump is applied along $\vec{x}$-direction.
% ---------------------------------------------------------
\begin{figure}[t]
\centerline{
~~~~~~
\includegraphics[width=8cm,height=6cm]{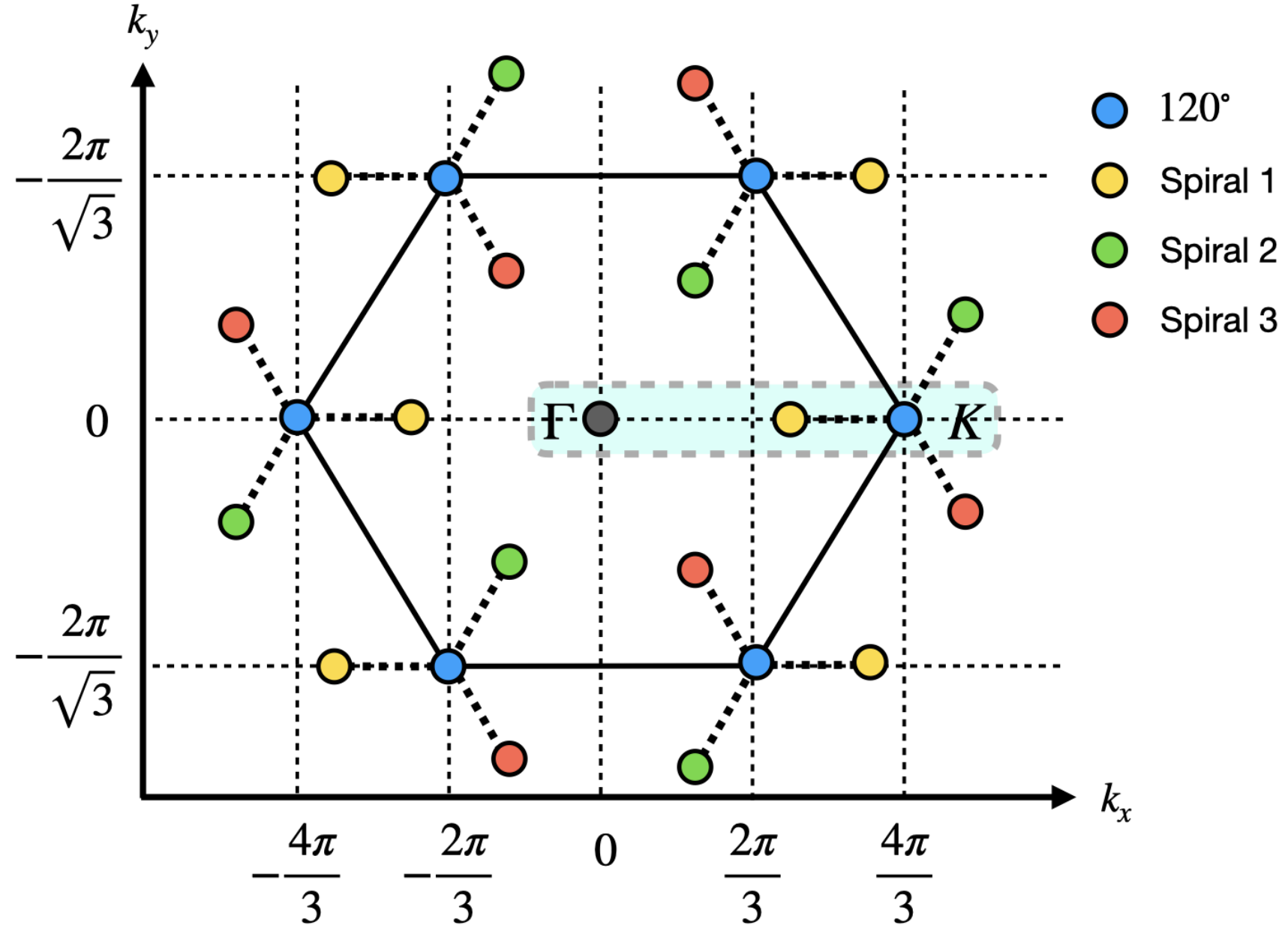}
}
\caption{Schematics of the spiral states. The Brillouin zone is 
shown with each corner (blues) depicting the $120^{\circ}$-order
in real space and the corresponding spiral states at incommensurate wavevector (yellow, green, red).
}
\end{figure}
% --------------------------------------------------------
% ---------------------------------------------------------
\begin{figure*}[ht]
\centerline{
\includegraphics[width=18cm,height=8.5cm]{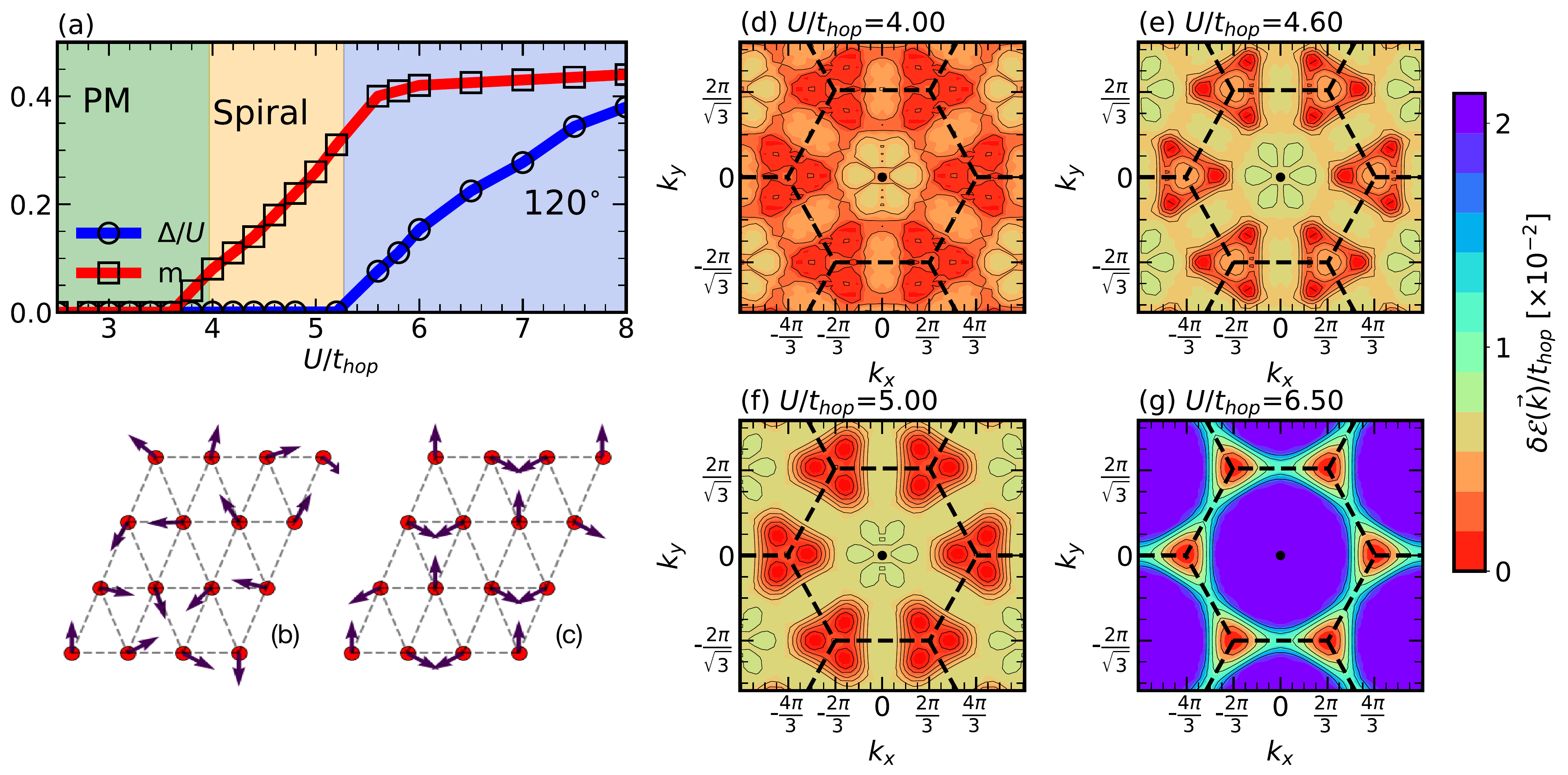}
}
\caption{
Ground state properties at equilibrium. 
(a) Variation of the magnetic moment and the gap in the density of states 
as a function of the interaction strength $U/t_{hop}$. In the 
mean-field approximation on a triangular lattice, the magnetic moment
begins to form at $U/t_{hop}=3.8$, and a gap $\Delta/U$ appears 
near $U/t_{hop}=5.2$. 
(b) A typical configuration of spiral order on a triangular lattice. 
Note that the unit cell for this magnetic order is larger than the 
$5\times 5$ lattice shown here. 
(c) The $120^\circ$-order with a unit cell of size $3 \times 3$. 
(d-g) The equilibrium energy landscape in the ordered space (see text 
for details). An instability occurs at a finite wavevector $\mathbf{Q}_1$, 
indicating a spiral state that progressively evolves towards the 
$\mathbf{Q}_0$ ($120^\circ$ order).
}
\end{figure*}
% ---------------------------------------------------------

\subsection{Indicators}

Our basic output from MFD 
is the time series for the equal-time 
density matrix $\rho_{ij}^{\sigma\sigma'}$ and from it
$\vec{m}_i(t)$.
Based on this we can compute various correlation functions of
the spin variables. 
Among these are the spatial transform ${\vec m}_{\bf q}(t)$,
the structure factor $S_{\bf q}(t)$, and the average
moment magnitude $m(t)$ defined below.
\begin{eqnarray}
\vec{m}_{\bf q}(t) &=& {1 \over N} \sum_{ij} e^{i {\bf q} 
\cdot {\bf R}_i } \vec{m}_i(t) \cr
 S_{\bf q}(t) &=&  |\vec{m}_{\bf q}(t)|^2 \cr
\cr
 m(t) & =& \frac{1}{N} \sum_i |\vec{m}_i(t)|
\end{eqnarray}

We also compute the
instantaneous electronic density of states (DOS) 
from the electronic eigenvalues $\epsilon_n$ in a
background $\vec{m}_i(t)$.  We can compute the 
`occupation' of these levels $\epsilon_n(t)$ from 
 $\rho_{ij}^{\sigma\sigma'}$.
The occupation function $P(\omega, t)$ for the instantaneous 
eigenstates, and the instantaneous DOS $\mathcal{N}(\omega,t)$ are
defined by:
\begin{eqnarray}
 P(\omega, t) \mathcal{N}(\omega,t) & = & \sum_n \rho_{nn}(t) \delta(\omega-\epsilon_n(t)) \cr
 \mathcal{N}(\omega, t) & = & {1 \over N} \sum_n \delta(\omega - \epsilon_n(t))
\end{eqnarray}
Where $\rho_{nn}$ is the expectation value of the
density operator associated with the $n$-th eigenstate with energy 
$\epsilon_{n}$: $\rho_{nn}(t) = 
\sum_{ij} U^*_{i\sigma, n}(t)U_{j\sigma',n}(t)\rho_{ij}^{\sigma \sigma'}(t)$, 
where $U(t)$ are the instantaneous eigenvectors.

\subsection{Ordering wavevectors}

In this paper, we deal with several states that occur at 
incommensurate wavevectors and some of them are related 
by symmetry. Here we briefly define how we denote
these
ordering wavevectors. In Fig.1. we schematically show the 
location of several ordering wavevectors that show up in
our results.  
Similar colors mean the same order in real space.
Each zone corner (blue) corresponds to $120^{\circ}$-order in 
real space. The line  connecting 
$\Gamma~(0,0)$ to  $K~(0,4\pi/3)$ is shaded light blue.
A point on this line is $(Q,0)$ (yellow circle). 
There are two more orders related to this and they are obtained by 
rotating the vector from $K$ to $Q$ by $120^{\circ}$
in (lattice) momentum space as shown in Fig.1. We call these 
Spiral 1, Spiral 2 and Spiral 3. In this paper, when we talk about the
Spiral state's wavevector ($Q$), for convenience it is usually 
about the wavevector of Spiral 1 
shown here on the $\Gamma-K$ line unless mentioned otherwise.

% ---------------------------------------------------------
\begin{figure*}[t]
\centerline{
\includegraphics[width=14cm,height=12cm]{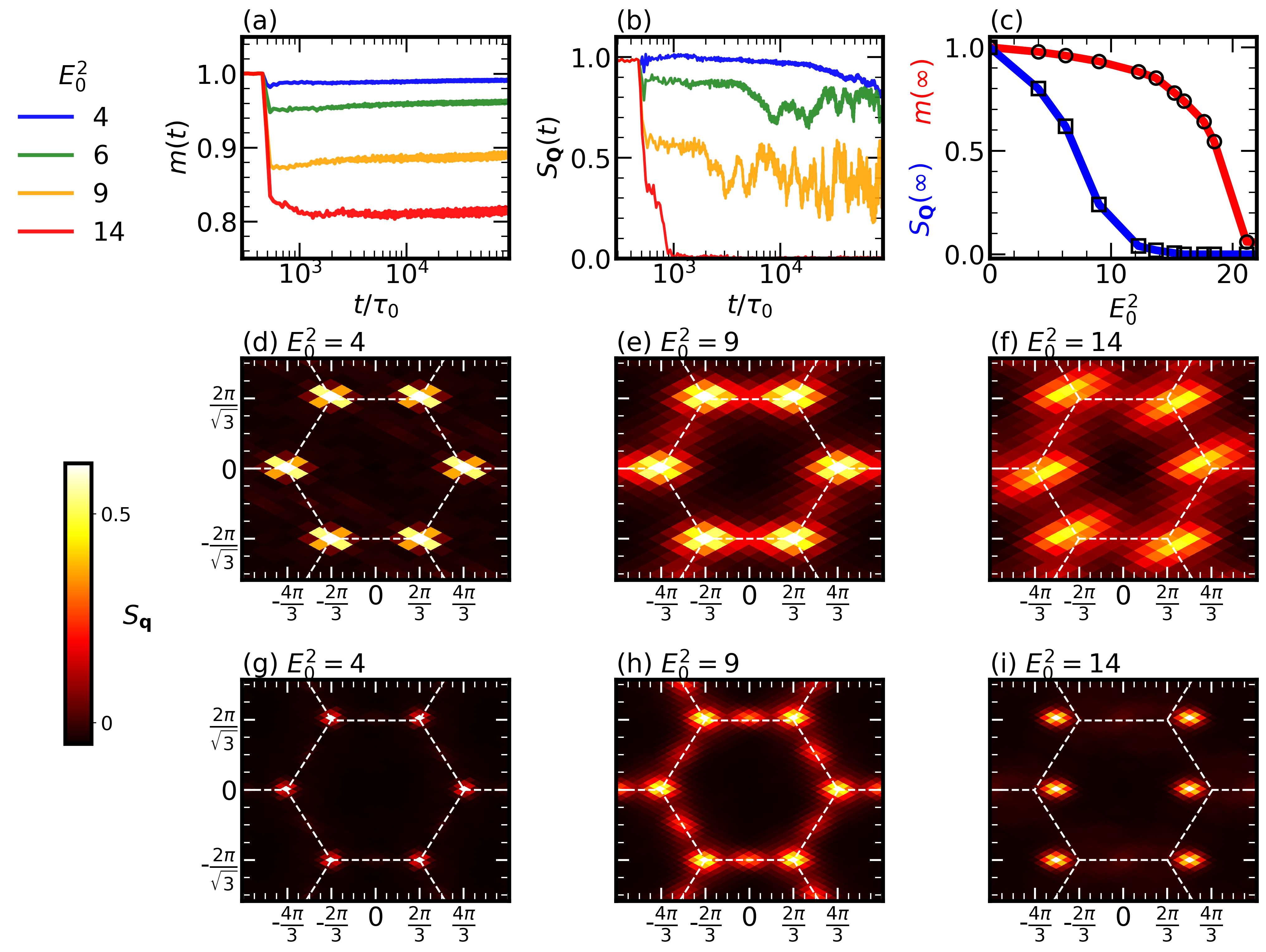}
~~~~~~~~~~~~~~~
}
\caption{
Effect of the pump pulse on the $120^\circ$ phase. 
(a)~Response of the average 
magnetic moment $m(t)$ to pulses with 
amplitude $E_0$. The pulse causes an `instantaneous' reduction 
in $m$, which remains constant thereafter to within 5\% variation. 
(b)~The structure factor $S_{{\bf Q}_0}$ associated
with $120^{\circ}$ order also has an immediate reduction similar 
to $m(t)$, followed by a slower long term decay to a steady value. 
(c)~Long-time values of $S_{{\bf Q}_0}$ and $m$ as functions of 
$E_0^2$. $S_{\bf Q}$ approaches zero around $E_0^2 = 12$, 
while $m$ goes to zero only around $E_0^2 = 21$. 
(d-f) The long time, $t \sim 10^5 \tau_0$, structure factor 
$S_{\bf q}$ over the full Brillouin zone on a 
$12\times 12$ lattice. At $E_0^2 = 4$, the weight is
spread around ${\bf Q}_0$, at $E_0^2 = 9$ the weight spreads
along the lines connecting the zone corners in momentum space, 
while at $E_0^2 = 14$ we get
$S_{{\bf Q}_0}\sim0$ with the suggestion of
a possible shift in the ordering wavevector.
(g-i)~Same results as above but on a $24\times 24$ lattice for 
$t \sim 5\times 10^3 \tau_0$. At
$E_0^2 = 14$ we now see a new ordering peak shifted away from
${\bf Q}_0$.
}

\end{figure*}
% ---------------------------------------------------------

\section{Ordering at Equilibrium}

Unlike the square lattice, where the $(\pi/a, \pi/a)$ state is always dominant, 
the Hubbard model at half-filling on a triangular lattice exhibits 
significant phase competition. In the ground state, the triangular 
lattice shows three distinct phases:
(i)~A correlated non-magnetic metal, referred to as PM, for 
\( U \leq U_{c1} \),
(ii)~A magnetic metal for \( U_{c1} < U < U_{c2} \), characterized 
by incommensurate spiral magnetic order with a wavevector \(\mathbf{Q}(U)\),
(iii)~A \(120^\circ\) ordered insulator for \( U \geq U_{c2} \).

The ground state phase diagram obtained from mean-field 
theory~\cite{tr7} predicts 
\( U_{c1}/t_{hop} = 0.66z t_{hop} \), where \( z \) is 
the coordination number (here \( z = 6 \)). The magnetic susceptibility 
shows a peak at wavevector \(\mathbf{Q}_1 = \mathbf{Q}(U_{c1}) = 
(0.73\pi/a, 0)\) or its symmetric wavevector. At \( U = U_{c1} \), 
local moment size \( m \) forms, and as \( U/t_{hop} \) 
increases \(\mathbf{Q}(U_{c1})\) gradually shifts from \(\mathbf{Q}_1\) 
to \(0.88\mathbf{Q}_0\) at \( U = U_{c2} \), where \(\mathbf{Q}_0 = 
(4\pi/3a, 0)\) represents a corner of the first Brillouin zone. 
We refer to this incommensurate phase as the `Spiral' state.
At \( U_{c2}/t_{hop} = 5.27 \), the ordering wavevector jumps
from \(0.88 \mathbf{Q}_0\) to \(\mathbf{Q}_0\) remaining 
constant with further 
increases in \( U/t_{hop} \), and the local moment
\( m \) jumps from 0.34 to 0.39 and asymptotically reaches the 
saturation value 0.5 as $U\rightarrow \infty$. 
In real space, this phase is characterized by a 
three-site sublattice antiferromagnetic order with a \(120^\circ\) 
twist between the sublattices, known as the \(120^\circ\)-order.

In Fig.2(a), we plot the magnetic moment size \( m \), which ranges 
from 0 to its saturation value of 0.5, starting from 
\( U_{c1}/t_{hop} \). The density of states gap \( \Delta \) 
opens at \( U_{c2}/t_{hop} \) and increases with \( U/t_{hop} \). 
A larger lattice numerical mean-field analysis shows the gap jumping 
from 0 to \(0.085~t_{hop}\) at \( U_{c2}/t_{hop} \). 
Fig.2(b) displays an example of 
the spiral state, with a wavelength larger than the lattice shown here, 
expected to repeat on a larger lattice. Fig.2(c) shows the \(120^\circ\) 
state.
Figs.2(d)-(g) present the mean-field equilibrium energy landscape 
\( \mathcal{E}(\vec{q}, U/t_{hop}) \) for various values of 
\( U/t_{hop} \). Just above \( U_{c1}/t_{hop} \), 
Fig.2(d) shows a weak energy well around \(\mathbf{Q}_1\). This 
energy well deepens with increasing \( U/t_{hop} \), but 
\(\mathbf{Q}(U/t_{hop})\) shifts towards \(\mathbf{Q}_0\) as
shown in Figs.2(e-f). Above \( U_{c2} \), the order is at 
\(\mathbf{Q}_0\), corresponding to the \(120^\circ\)-order in real
space, as shown for \( U/t_{hop} = 6.4 \) in Fig.2(g). 
In this study, we primarily present our dynamic results at 
\( U/t_{hop} = 6.4 \), where \( m \sim 0.42 \) and 
the gap \( \Delta/t_{hop} \sim 1.3 \).
%----------------------------------------------------
\begin{figure*}[t]
\centerline{
~~~~~~\includegraphics[width=11.5cm,height=3.5cm]{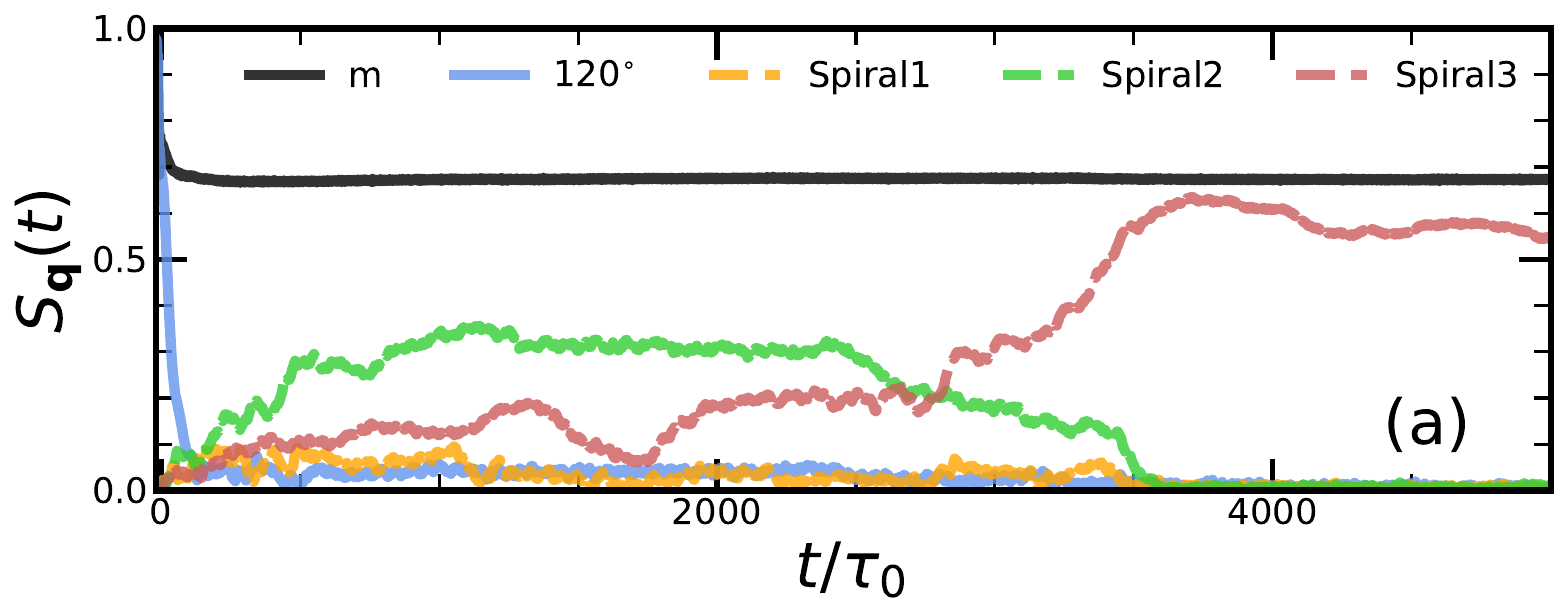} }
\centerline{
\includegraphics[width=11.3cm,height=3.8cm]{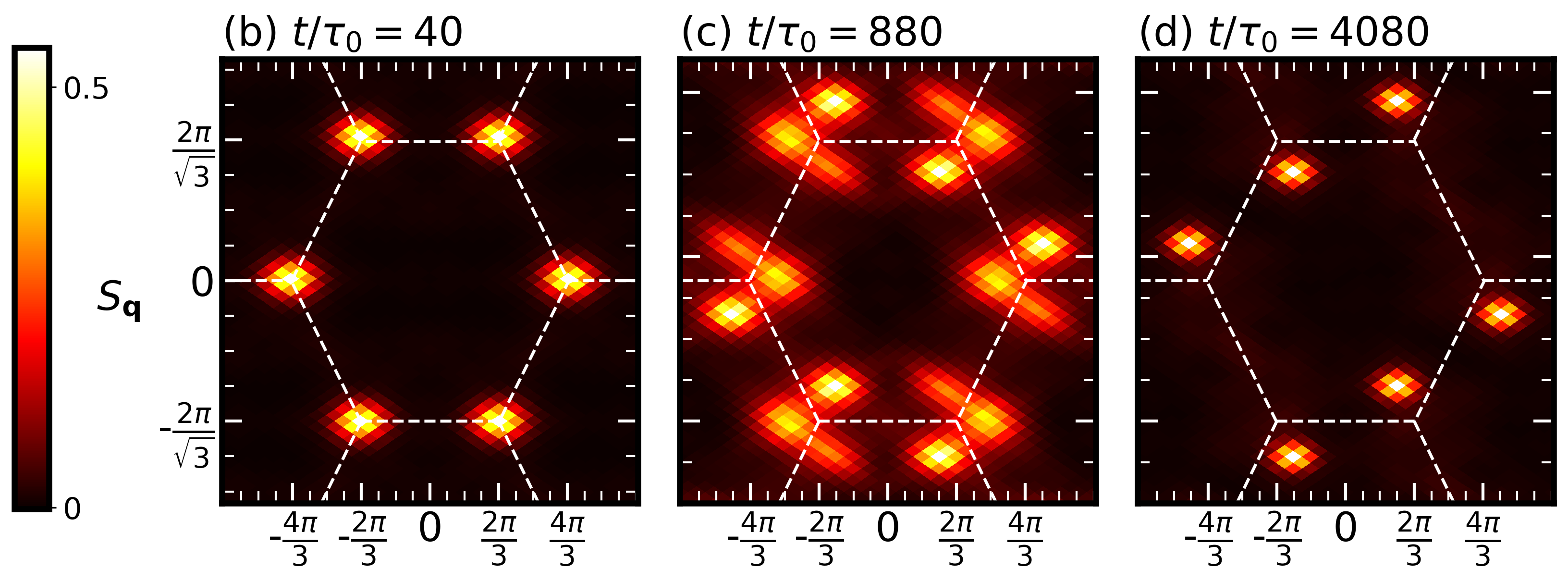} }
\centerline{
\includegraphics[width=11.3cm,height=3.8cm]{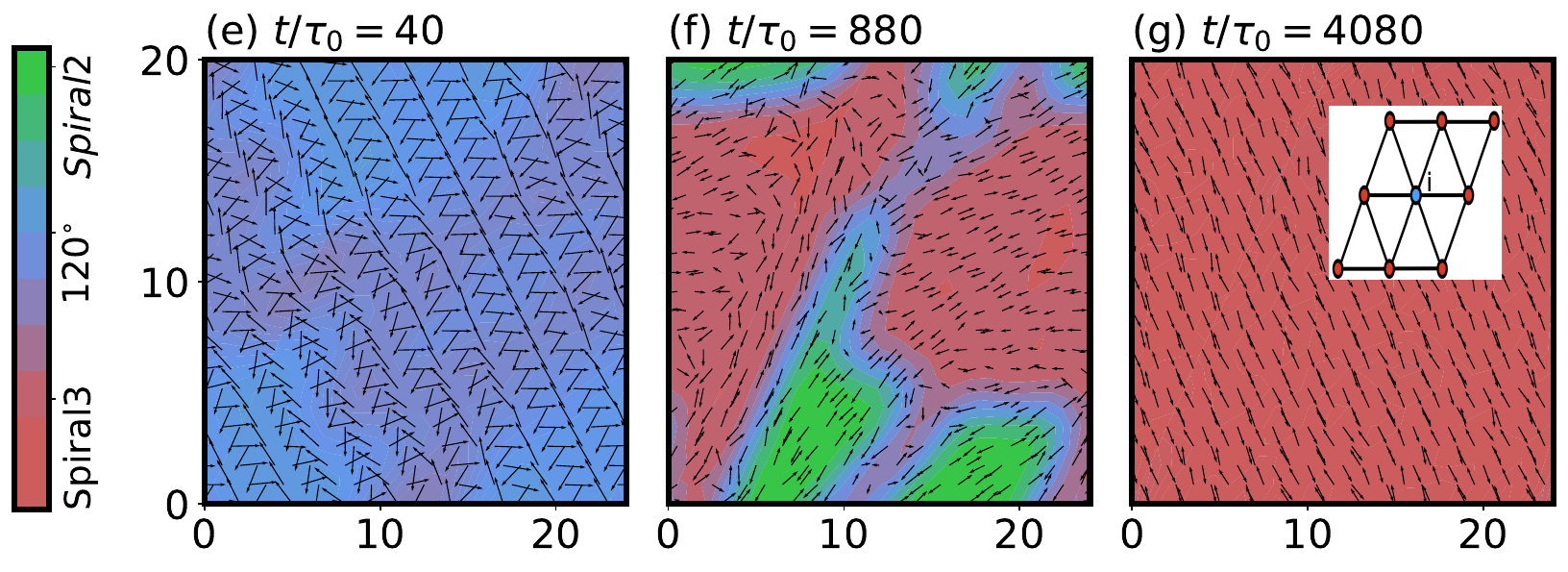} }
\centerline{
~~~~~\includegraphics[width=11.0cm,height=3.5cm]{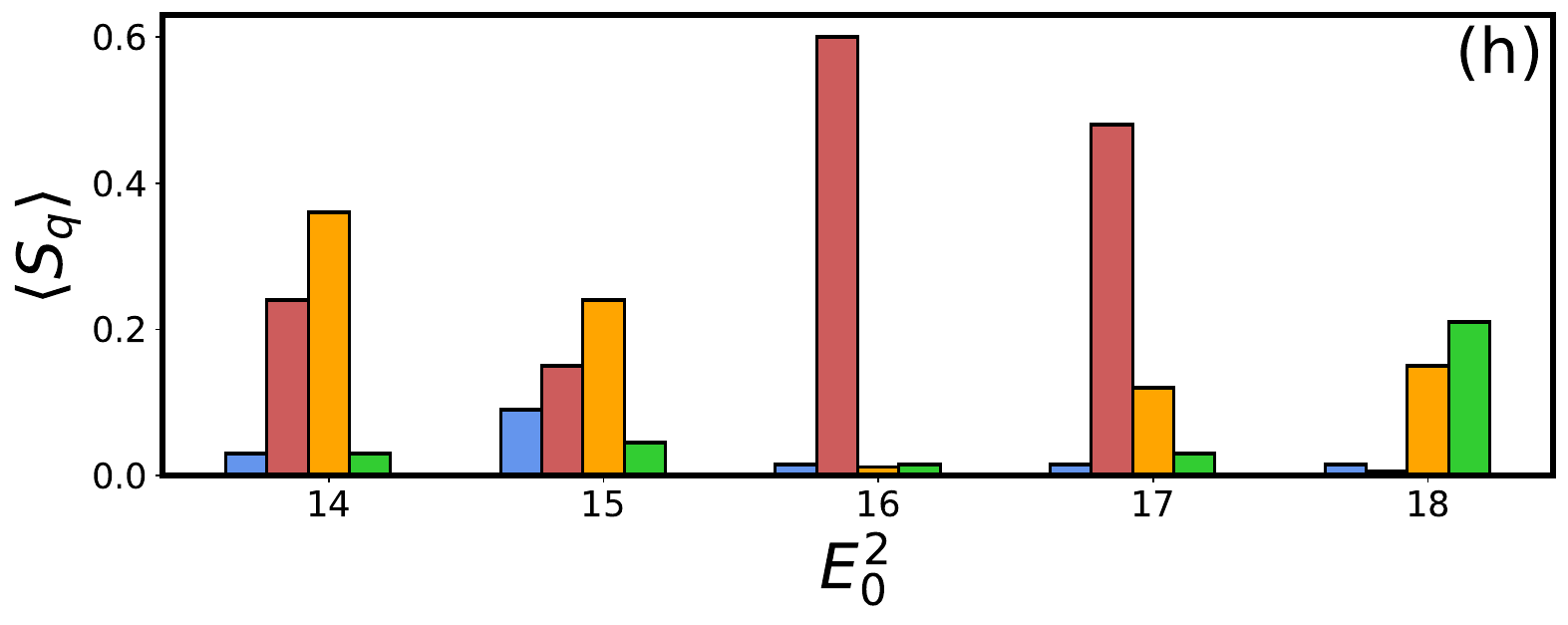} }
\caption{
Dynamics of domain formation at $E_0^2 = 16$.
(a) Time evolution of structure factors associated with $120^\circ$ 
order ($S_{\mathbf{Q}}(t)$, blue) and various spiral orders 
(yellow, green, red). The decay of $S_{\mathbf{Q}_0}(t)$ within 
a timescale of $t/\tau_0 = 100$ is followed by the emergence 
of spiral orders, whose domains compete until $t \sim 3500 \tau_0$. 
After this, one spiral phase dominates. 
(b)-(d) Color maps of $S_{\mathbf{q}}$ at different times show a 
transition from sharp Bragg peaks at zone corners around $40 \tau_0$ 
to a broad distribution around $880 \tau_0$, culminating in a 
distinct peak at the spiral wavevector at $4080 \tau_0$. 
(e)-(g) Local order (see text), with blue indicating the 
$120^\circ$ order. Around $880 \tau_0$, domains of Spiral 3 
and Spiral 2 orders are prominent, with Spiral 3 ultimately 
prevailing.
(h) shows the proportion of different orders at 
$t/\tau_0=4\times 10^3$.
}
\end{figure*}
% ---------------------------------------------------------
\section{Results from mean field dynamics}
\subsection{Order Parameter Dynamics}
At \( U/t_{hop} = 6.4 \), the ground state of the triangular 
lattice is an insulator with \(120^\circ\) magnetic order. We 
initialize the system with an ordered state corresponding to 
wavevector \(\mathbf{Q}_0\), 
and introduce a small fluctuation over the order state to prepare 
an initial thermal configuration at temperature \(T = 10^{-6} t_{hop}\). 
We use a \(12 \times 12\) lattice to study the long-time behavior 
of the mean magnetic moment \( m(t) \) and the structure factor 
at the ordering peak \( S_{\mathbf{Q}_0}(t) \).

A laser pulse rapidly reduces the magnetic moment \( m \) 
to a value that depends on the pulse amplitude \( E_0 \), 
after which it stabilizes. In Fig. 3(a), we observe that 
the magnetic moment \( m(t) \) after the pulse can 
decrease to approximately 80\% of its original value 
for \( E_0^2 \sim 12 \), while the dominant 
\( 120^\circ \) order remains unchanged. 
The structure factor \( S_{\mathbf{Q}_0}(t) \) shows 
a three stage dynamics with
an initial sharp decline, followed by a slow decline and then 
followed by oscillations around a steady mean, as shown in Fig. 3(b).

At \( E_0^2 \sim 12 \), the long-time 
value of \( S_{\mathbf{Q}_0}(t \rightarrow \infty) \) approaches zero, 
as shown in Fig.3(c). The mean magnetic moment
\( m(t \rightarrow \infty) \) however 
remains about \(80\%\) of its 
initial value. This indicates that the loss of \(120^\circ\) 
order is not due to the quenching of magnetic moments.
Beyond pump strength ${E_0}^2\sim 21$ the 
\( m(t \rightarrow \infty) \) goes to $5\%$ of its initial value
with no long range order.

We plot the map of the full structure factor 
\( S_{\mathbf{q}}(t \rightarrow \infty, E_0) \) 
for the \(12 \times 12\) system in the middle 
row  of Fig.3. For a low pump amplitude (\( E_0^2 = 4 \)),
Fig.3(d) shows a broad distribution around the zone corners, 
including \(\mathbf{q} = \mathbf{Q}_0\). For a higher pump 
amplitude (\( E_0^2 = 9 \)), Fig.3(e) reveals that the weight
distributes along the zone boundary. At a pulse strength 
\( E_0^2 = 14 \), which exceeds the critical value \( (E_0^{c1})^2 = 12 \), 
the order appears to shift towards a different wavevector. However, 
due to the limited resolution (\(\delta \mathbf{q} \sim 4\pi/3\sqrt{N}\)) 
in the small system, capturing this shift is challenging.

We extended our study to a larger \(24 \times 24\) system to 
address this.
As the computation cost increases, we can simulate up 
to \( \tau_{max} \sim 5\times 10^3 \tau_0 \), which is an order 
of magnitude smaller than the $\tau_{max}$ for the 
\(12 \times 12\) lattice. The structure factor 
\( S_{\mathbf{Q}_0}(t \rightarrow \infty) \) for the \(24 \times 24\) 
system is shown in the bottom row  of Fig.3. It confirms 
the basic features, seen in the smaller system, and shows
that at $E_0^2 = 14$ there is a new ordering peak in the
Brillouin zone.
 
In Fig.4, we present a detailed analysis of the magnetic configurations 
on a \(24 \times 24\) lattice at a pump strength of \( E_0^2 = 16 \). 
Fig.4(a) shows the time evolution of the mean magnetic moment 
\( m(t) \) and the structure factor at wavevectors corresponding 
to \(120^\circ\) order and various spiral states (Spiral1, Spiral2, 
and Spiral3). The magnetization \( m(t) \) drops quickly
to approximately \(0.7\) and then stabilizes. The \(120^\circ\) 
order decays within \( t = 100 \tau_0 \), giving way to competition 
between Spiral2 and Spiral3 orders. Eventually, around \( t = 3000 \tau_0 \), 
the domain associated with Spiral3 order becomes comparable to the 
system size.

The structure factor \( S_{\bf q} \) is displayed in the second 
panel, Fig.4(b)-(d). At \( t = 40 \tau_0 \), the initial weight is 
concentrated at \( \mathbf{q} = \mathbf{Q} \). By \( t = 880 \tau_0 \),
\( S_{\mathbf{q}} \) shows a broad distribution around \( \mathbf{Q}_0 \),
indicating a lack of long-range order. Subsequently, \( S_{\mathbf{q}} \) 
becomes concentrated around a spiral order by \( t = 4080 \tau_0 \).
% ---------------------------------------------------------
\begin{figure}[b]
\centerline{
\includegraphics[width=8.5cm,height=7.5cm]{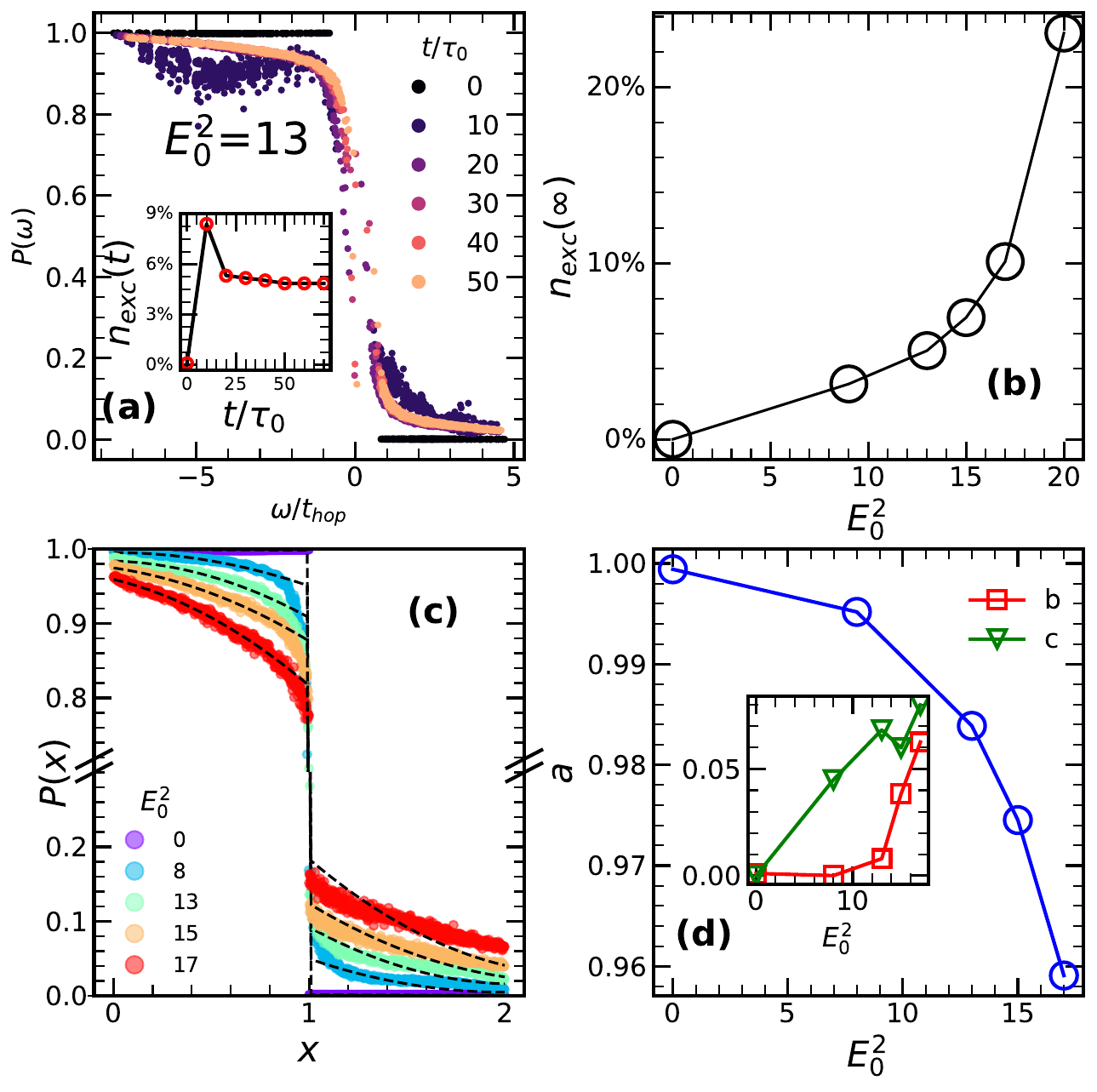}
}
\caption{
Electronic occupation function.
Top panel: (a) Time evolution of the occupation 
function $P(\omega)$ at $E_0^2 = 13$ as calculated from 
MFD. Pre-pulse $(t=0)$ the function $P(\omega) = \theta(-\omega)$.
Due to the pulse a non trivial $P(\omega)$ is created at short
times, with an associated $n_{exc}(t)$.
 $P(\omega)$ settles into a
steady-state within $t \sim 50 \tau_0$. The inset shows 
upper band population $n_{exc}(t)$
rising from 0 to $\sim 9\%$ and then stabilizing
around $\sim 6\%$. The rapid convergence to a steady state is
similar for other $E_0^2$ values up to $E_0^2 \sim 20$.
(b) Steady-state value of $n_{exc}$ as a function of $E_0^2$.
Bottom panel:
(c) Occupation function plotted with respect to the 
normalised eigenstate number $x=n/N$, where
$N$ is the total number of sites.
$x$ ranges from 0 to 2. $P(x)$ at long time for various 
$E_0^2$ is fitted with the function described in the text.
(d) Dependence of the fitting parameters $a(E_0^2)$
(and $b(E_0^2)$ and $c(E_0^2)$, shown in the inset).
}
\end{figure}
% ---------------------------------------------------------

% ---------------------------------------------------------
\begin{figure*}[t]
\centerline{
\includegraphics[width=8.8cm,height=8cm]{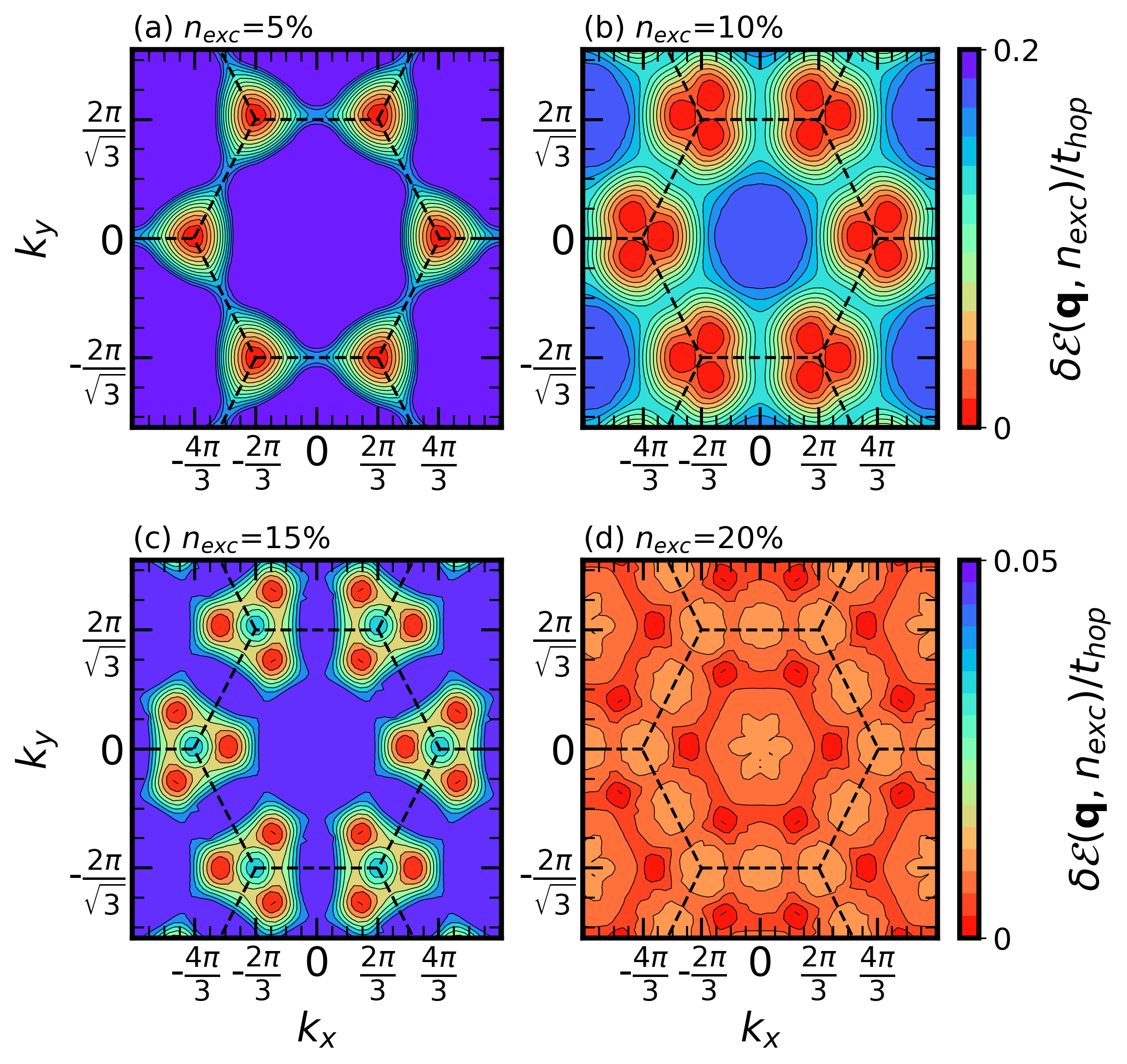} 
\includegraphics[width=8.4cm,height=7.5cm]{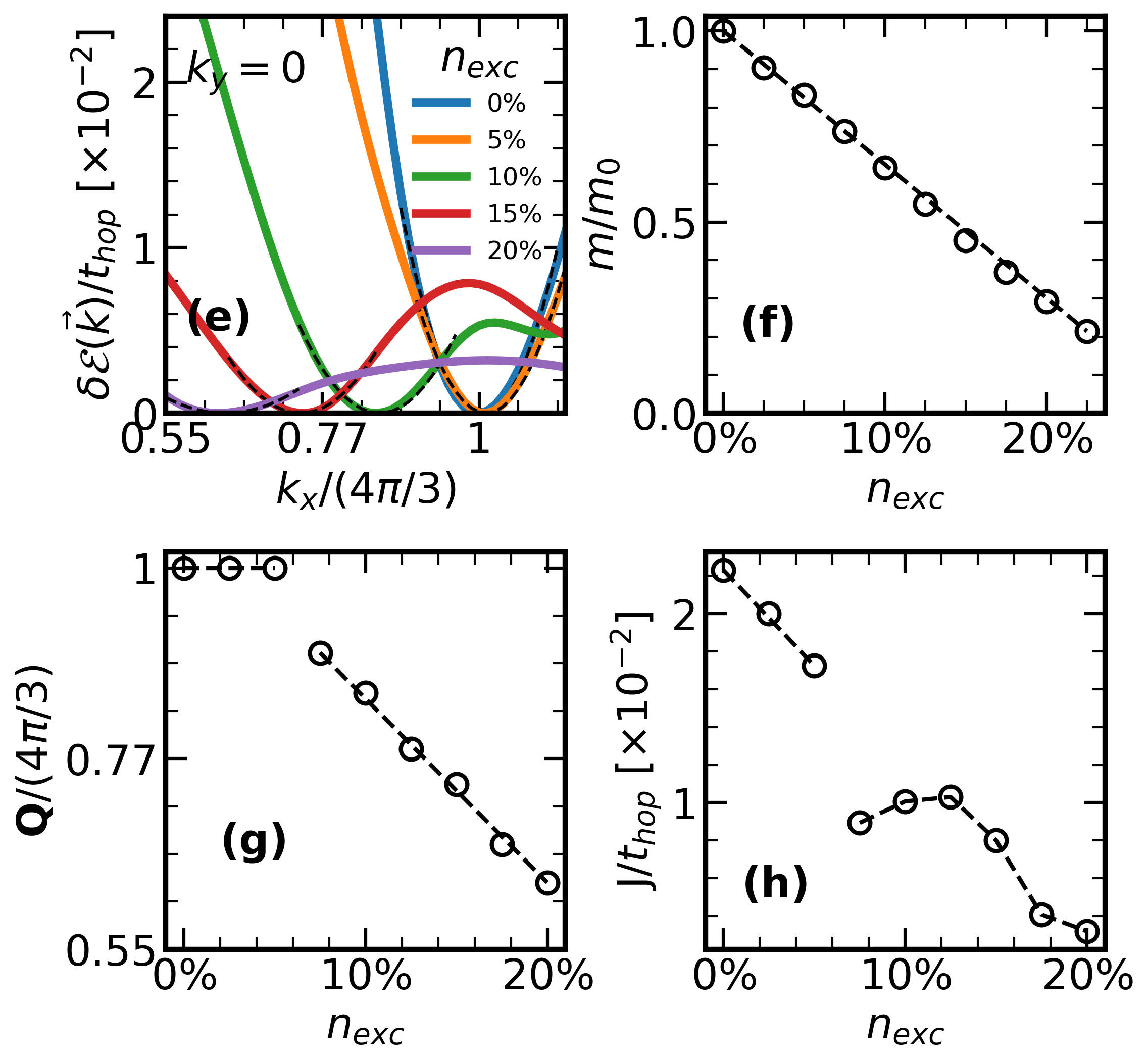} 
}
\caption{
Nonequilibrium energy landscape at $U/t_{hop} = 6.4$.
(a) For low \(n_{exc}\), the minimum is at ${\bf Q}_0$.
(b) At \(n_{exc} \sim 10\%\) 
the minima shift to spiral states.
(c) New minima pronounced at 
\(n_{exc} \sim 15\%\). 
(d) When \(n_{exc} \sim 20\%\), the 
local moment magnitude is very small,
leading to shallow energy minima suggesting small 
thermal fluctuations can give rise to a paramagnetic phase with
weak local-moments. 
(e) The energy \(\delta \mathcal{E}(\vec{q})/t_{hop}\), with
respect to the minimum, 
for $q_x$ along \((0,0) \rightarrow (4\pi/3,0)\), showing 
how the location of the minimum changes with \(n_{exc}\).
The dotted line shows the fit of \(\frac{1}{2}J (q-Q)^2\) 
around \(q=Q\). 
(f) The magnitude of \(m\) at the minima of the energy 
landscape is plotted as a function of \(n_{exc}\).
(g) The ordering wavevector \(Q\) as a function of 
\(n_{exc}\) shows that the \(120^\circ\) order 
(with \(Q_0 = 4\pi/3\)) remains stable up to \(n_{exc} 
\approx 5\%\). 
Beyond this point, \(Q\) drops sharply from \(Q_0\) to 
approximately \(0.9~Q_0\) and then decreases linearly 
with increasing \(n_{exc}\).
(h) The effective 
spin compressibility
 \(J/t_{hop}\) decreases 
linearly up to \(n_{exc} \sim 5\%\) from \(0.022\) to \(0.016\). 
It then shows a discontinuous jump to \(0.008\) at \(n_{exc} 
\sim 7.5\%\), increases to \(0.01\) at \(n_{exc} \sim 15\%\), 
and decays to \(\sim 0\) around \(n_{exc} \sim 20\%\).
}

\end{figure*}
% --------------------------------------------------------
Panels Fig.4(e)-(g) show snapshots of the real-space 
configurations. 
We plot the colormap of the local correlation 
$ C_i = \sum_{j \in \mathcal{J}} \vec{m}_i.\vec{m}_j$,
where \(\mathcal{J}\) includes all 6 nearest neighbors and two 
opposite 
next-nearest neighbors, as depicted in the inset of Fig.4(g). This 
measure breaks the symmetry between the individual spiral states.
Figs.4(e)-(g) use color coding to represent different orders: 
\(120^\circ\) (blue), Spiral2 (green), and Spiral3 (red). 
Notably, no particular spiral state is favored when the pump 
strength exceeds
\( E_0^c \). 
Fig.4(h) shows the
proportion of different orders as a function of \( E_0^2 > 
\left(E_0^{c1}\right)^2 \) at $t/\tau_{0}\sim 4\times 10^3$ which is 
equivalent to the largest simulation time.
For \( E_0^2 > 21 \), the magnetic moments become very small (\( m 
\sim 0.04 \)) and no long-range order is observed.\\

~~Our simulation time, $\tau_{\text{sim}}$, is $5 \times 10^3 \tau_0$. 
Within this period, we observe the following results in the 
intermediate pumping regime ($E_0^2 \approx 12$ to $20$):
(i) The domain nucleation process is relatively fast, occurring 
on the order of $100 \tau_0$.
(ii) Minor changes in initial conditions can lead to the 
competitive emergence of various spiral orders.
(iii) Domain growth is slow and depends on the pump strength 
$E_0^2$.
(iv) Near the edges of this regime, different spiral orders 
continue to compete at $t = \tau_{\text{sim}}$.\\
~~Although MFD reveals these new ordered 
states, determining the domain growth timescale is challenging 
due to the limitations of numerical simulation resources. 
This suggests that a longer simulation time is required, 
which may depend on $E_0^2$ and could potentially diverge. 
Additionally, averaging results over initial states in MFD 
calculations is necessary.
In the following section, we will explain how an energy 
landscape-based method can provide insights into the formation 
of new spiral states. We will also use a Langevin 
dynamics-based approach, where the simulation time can be up 
to $10^3$ times longer than that in MFD for a similar
$24 \times 24$ system at similar numerical cost, 
to demonstrate how domain formation 
time might depend on pump strength.

\subsection{Electronic population}

Prior to the pulse, the electronic population follows a Fermi function at 
zero temperature, indicating a fully filled lower Hubbard band. Upon 
introducing the pump, a fraction of the electrons transition to the 
upper Hubbard band. Despite the excited population and the 
associated suppression of the magnetic moment, a gap remains in the 
density of states upto $E_0^2 \sim 20$.

The population \(P(\omega)\),
calculated as described in Section II, is
shown  for \(E_0^2 = 13\) and 
different times in Fig.5(a).
By $t \sim 50 \tau_0$  \(P(\omega)\) assumes its long time
form. We have found that the time to attain this `steady state'
form does not depend significantly on $E_0$.  
The population in the upper band, 
\(n_{exc}(t) = \int_{0}^{\infty} d\omega \, N(\omega,t)P(\omega,t)\), 
is plotted in the inset of Fig.5(a). Following the
pulse \(n_{exc}\) 
rises to approximately 9\% and then reduces to
reach a steady-state value of around 
5\%. The steady-state value of 
\(n_{exc}\) varies with pump strength \(E_0^2\) 
as shown in Fig.5(b).

The bottom panels of Fig.5 characterize the nonequilibrium
population function in the steady state. We describe the 
occupation 
function using instantaneous eigenstate numbers \(x\), where 
\(x = 1\) represents half-filling, with \(x\) ranging from 0 
to 2. We fit this occupation function with:

\begin{equation}
    P(x) = \begin{cases} 
    a - b x - c x^2 & \text{for } x < 1, \\
    1 - \left(a - b (2 - x) - c (2 - x)^2\right) & \text{for } x \geq 1.
    \end{cases}
\end{equation}

% -----------------------------------------------------
\begin{figure*}[t]
\centerline{
\includegraphics[width=14cm,height=4.2cm]{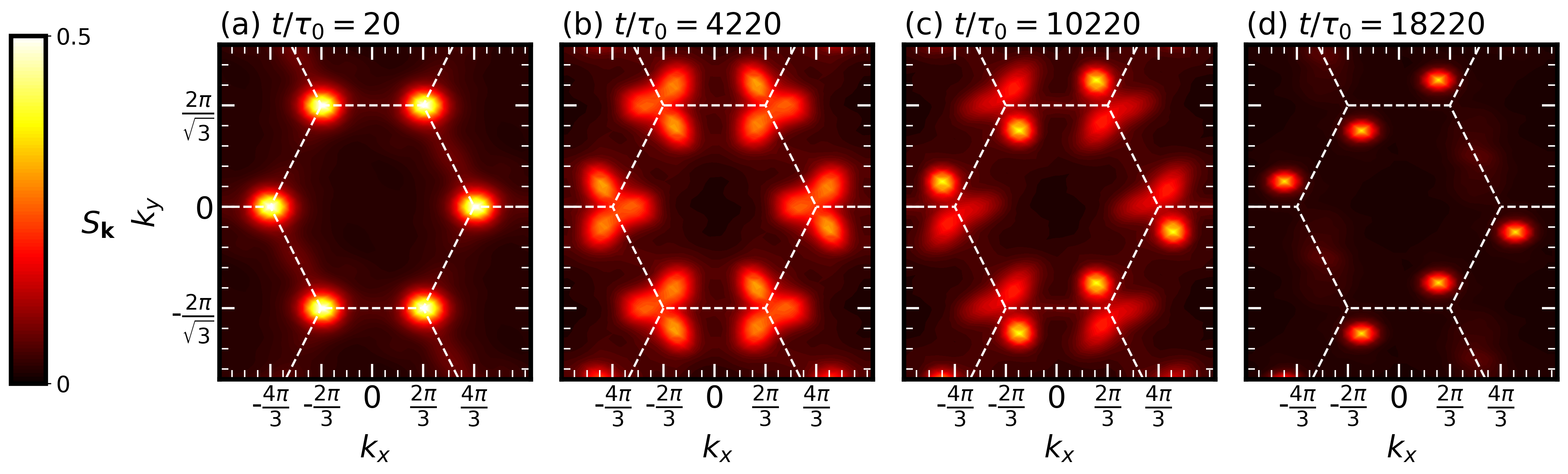}
}
\centerline{
~~~~~~~~~~~~~~~
\includegraphics[width=6.5cm,height=4.5cm]{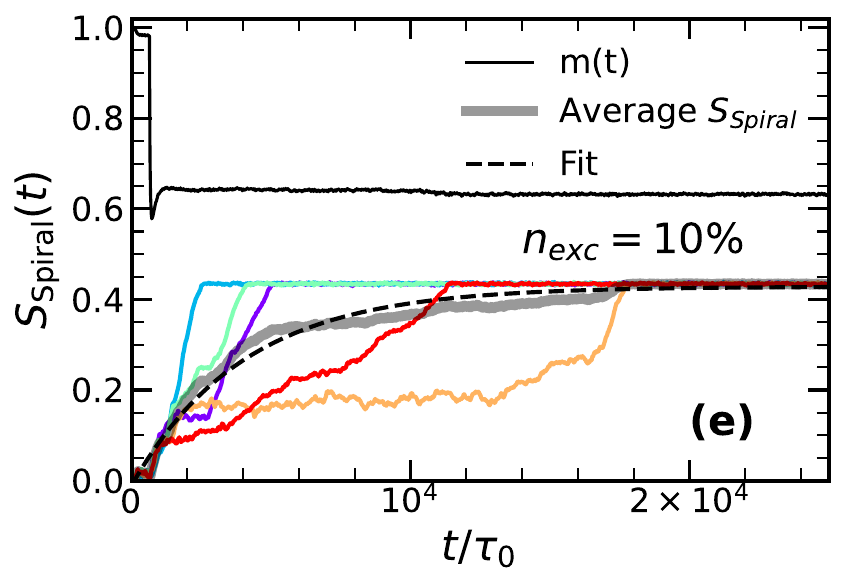}
\includegraphics[width=6.5cm,height=4.5cm]{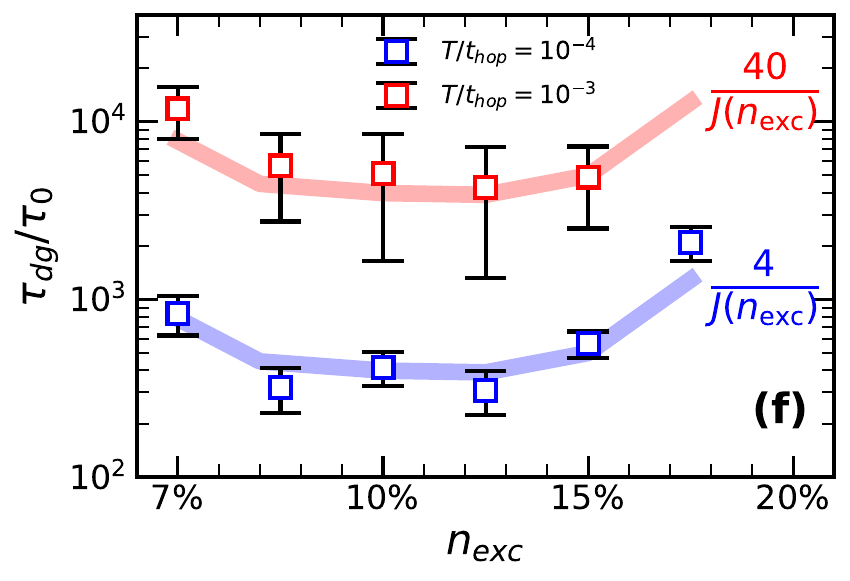}
}
\caption{
Domain growth from Langevin dynamics.
(a-d) The maps of the structure factor $S_{\mathbf{q}}(t)$ over the 
full Brillouin zone with $n_{exc} = 10\%$ at a temperature 
of $T/t_{hop} = 10^{-3}$. At $t/\tau_0 = 20$, the map shows 
weight at $\mathbf{Q}_0$. By $t/\tau_0 = 4\times 10^3$, broad 
weight is observed around $\mathbf{Q}$. At $t/\tau_0 = 10^4
$, the broad weight begins to condense around the spiral 
wavevector $\mathbf{Q}^*$. Finally, at $t/\tau_0 = 1.8\times 10^4$, 
long-range order at $\mathbf{Q}$ emerges.
(e) The time evolution of $m(t)$ and different stochastic traces 
for the spiral order indicator $S_{Sp}(t)$ are shown, 
with their average depicted. Different colors represent various 
thermal runs. The average, indicated by the faint black line, is 
fitted with the function 
\( S_{sp}(t) = S_0 (1 - e^{-t/\tau_{\text{dg}}}) \). 
This analysis was conducted on a $24 \times 24$ lattice with 
a simulation time window of $\tau_{max} = 10^6\tau_0$.
(f) The dependence of $\tau_{dg}$ on $n_{exc}$
and temperature $T$ is illustrated. The faint solid lines show
that $\tau_{dg}$ behaves roughly as $\propto 1/J(n_{ex})$.}
\end{figure*}
% -----------------------------------------------------
% ---------------------------------------------------------
\begin{figure}[b]
\centerline{
\includegraphics[width=4cm,height=3.5cm]{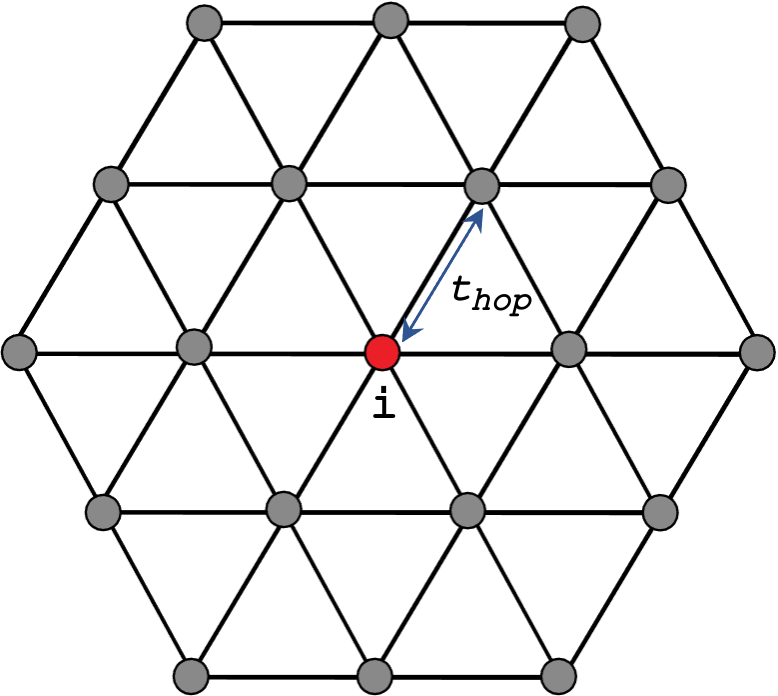}
}
\caption{The cluster centered on ${\bf R}_i$ that is used to
calculate the torque on ${\vec m}_i$ by diagonalising $H_{SF}$. 
Apart from the central site, it has $6$ nearest neighbours and
$12$ next nearest neighbours. 
}
\end{figure}
% --------------------------------------------------------
This fit ensures that the total electron count remains constant, as 
is the case in mean-field dynamics (MFD). The fits 
for various \(E_0^2\) values are shown in Fig.5(c). The detailed
dependence of the parameters \(a\), \(b\), and \(c\) on \(E_0\) is
presented in Fig.5(d). Despite its simplicity, this fitting 
function effectively captures the steady-state population across
 different \(E_0^2\) values with just a few parameters.

\section{Nonequilibrium energy landscape}

The emergence of new long range orders due to
pump excitation suggests a modification
in the energy landscape. We examine the energy 
\(\mathcal{E}(m, q_x, q_y)\) associated with variational 
states exhibiting long-range order at wavevector \((q_x, q_y)\) 
for finite \(n_{exc}\). These landscapes are characterized 
by a unique \(n_{exc}\) value. As illustrated in Fig.6(a)-(d), 
for \(n_{exc} < 5\%\), the minimum energy is for \(120^\circ\)
order. 
Increasing \(n_{exc}\) alters the energy landscape, indicating 
the onset of spiral orders in the range \(6\% < n_{exc} < 20\%\).
This range reveals three symmetric \(\vec{Q}\)-points as 
minima of the new landscape. With increasing excitation, the 
wavevector \(\vec{Q}\) shifts, and the depth of the minimum decreases. 
For \(n_{exc} > 20\%\), the energy landscape becomes nearly 
flat, suggesting a transition to a paramagnetic phase
with a small moment size. 

Low-energy features of the energy landscape can be characterized by 
fitting a harmonic potential around the minima (at any of new 
wavevectors \(\vec{Q}\)) as \(V(\delta q)=\frac{1}{2} 
J |\vec{\delta q}|^2\), 
where \(\vec{\delta q} = \vec{q} - \vec{Q}\). Fig.6(e) shows 
\(\delta \mathcal{E}(\vec{k})/t_{hop}\) along \((0,0) 
\rightarrow (4\pi/3,0)\), illustrating how the location of the
minima \(\mathbf{Q}\) changes with \(n_{exc}\). The 
dotted line represents the fit of \(\frac{1}{2}J (q-Q)^2\) 
around \(q=Q\). Fig.6(f) plots the magnitude of \(m\) at the 
minima of the energy landscape as a function of \(n_{exc}\), 
showing a linear suppression of \(m\). Fig.6(g) demonstrates that the
ordering wavevector \(Q\) is stable 
at (\(Q_0 = 4\pi/3\)) up to \(n_{exc} \sim 5\%\). 
Beyond this, it shifts from \(4\pi/3\) to approximately
\(0.9 \times 4\pi/3\) and decreases linearly with increasing 
\(n_{exc}\). Fig.6(h) shows that the effective compressibility 
\(J/t_{hop}\) decreases linearly up to \(n_{exc} \sim 5\%\), 
from \(0.022\) to \(0.016\). It then experiences a discontinuous jump to 
\(0.008\) at \(n_{exc} \sim 7.5\%\), increases to \(0.01\) at 
\(n_{exc} \sim 15\%\), and decays to approximately \(0\) around 
\(n_{exc} \sim 20\%\).

This analysis suggests the possibility of an order-to-order magnetic 
phase transition due to pumping when the bath temperature \(T\) is 
smaller than \(J\). This transition involves long-range order, and 
over long timescales, the system should evolve towards these states 
if thermal fluctuations are sufficiently small and the nonequilibrium 
population does not decay within this timescale. However, 
detailed dynamics from an initial \(120^\circ\) order to a 
spiral state require further investigation. Probing these dynamics
is computationally intensive with full mean-field dynamics (MFD).

\section{Langevin dynamics for domain growth}

We utilize the Langevin scheme, as defined previously, to 
directly explore 
the dynamics of \(\vec{m}_i\). We calculate the torque 
on \(\vec{m}_i\) by incorporating the effect of \(n_{exc}\) 
on the change in 
electronic energy due to an incremental change in \(\vec{m}_i\). 
This in principle requires diagonalizing the entire system to obtain 
the eigenvalues and then using the population function.
For large \(U\), however, one can calculate the torque 
by using 
a smaller cluster around site \(i\) and diagonalizing the Hamiltonian 
only on this smaller cluster. 
We find that a cluster up to the 
next nearest neighbors, comprising 19 sites, is sufficient to 
capture the equilibrium properties. 

The model can now be simulated over a longer timescale 
(\(\sim 10^6\tau_0\)) on larger system sizes to capture domain 
dynamics. Using a \(24 \times 24\) lattice, we examine the domain
growth process and determine the timescale for domain growth as a 
function of \(n_{exc}\) and the bath temperature 
\(T/t_{hop}\).
In absence of any external thermal 
bath, the thermal fluctuations 
comes from integrating out the excited electrons.
In reality, bath temperature $T$ and dissipation rate $\gamma$
should be derived from the microscopics 
of the system which is beyond the scope of this study.
We treat these parameters as phenomenological. 
Though in realistic situations 
presence of an external bath has to be considered as well as 
dissipation coming from phonons, it should be pointed out that  
this Langevin scheme can be generalized to accommodate those 
situations. 

In Fig.7(a)-(d), we plot the map of \(S_{\bf q}\) at \(n_{exc} = 10\%\) 
across the entire momentum space. Long-range order emerges around
\(t/\tau_0 \sim 1.8 \times 10^4\), which is significantly longer 
than achievable with MFD. There are three possible spiral states 
that are degenerate, and the long-time domain can be any 
one of these states. To account for the stochastic nature of the 
process, we run 5 to 6 thermal simulations for each evolution.
The solid black line in Fig.7(e) represents the average magnetic 
moment \(m(t)\), showing a rapid reduction from 1 to \(\sim 0.6\). 
Different colors in \(S_{\bf q}(t)\) indicate various thermal runs. 
We average these thermal runs, shown by the faint black line, and 
fit this average with an exponential growth function 
\(S_{\bf Q}(t) = S_0 (1 - e^{-t/\tau_{dg}})\), where \(\tau_{dg}\) 
is the timescale required to grow to the system size. This allows 
us to extract \(\tau_{dg}\) for different \(n_{exc}\) values. 
Fig.7(f) illustrates that \(\tau_{dg}\) grows rapidly at two points: 
around \(n_{exc} = 7\%\), where the transition from the
\(120^\circ\) state to spiral states occurs, and around 
\(n_{exc} = 20\%\), where \(m\) becomes very small. 
This roughly behaves like $\sim 1/J(n_{exc})$ 
as shown with faint solid-lines in Fig.7(h). 
The error bars in Fig.8(f) represent 
the one-sigma width of these averages, which are substantial. 
As the temperature \(T/t_{hop} = 10^{-3}\) increases, 
the domain growth timescale also increases by an order of 
magnitude, as indicated in red.

A typical cluster that is used for calculating the torque on
a local moment is shown
in Fig.8. Using this the system update
cost within LD becomes 
\(\mathcal{O}(N)\).

\section{Discussion}

\subsection{Nonequilibrium phase diagram}

The nonequilibrium population induced by the pump pulse 
depends on several factors, including the pumping frequency 
\(\omega_{p}\), electric field amplitude \(\vec{E}_0\), 
pump width \(\tau_{p}\), and the system parameters. 
However one can create a `phase diagram' using $n_{exc}$
itself as an input, without worrying about the specific combination of $\omega_p,~\tau_p,~E_0$ from which it arises. In that spirit we construct a nonequilibrium phase diagram using
the population function constructed earlier, parametrised
by $n_{exc}$ and the $U/t_{hop}$ value. This is based on
simple minimisation as set out for our variational
calculation in Section II.

% ---------------------------------------------------------
\begin{figure}[t]
\centerline{
~~
\includegraphics[width=8cm,height=5.5cm]{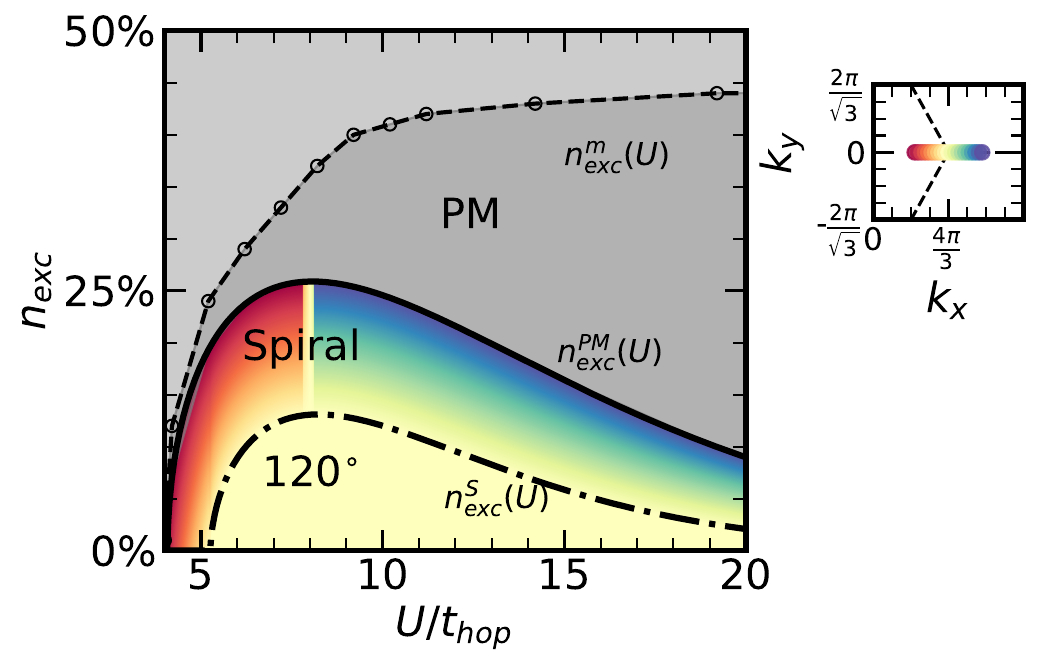}
}
\caption{
Nonequilibrium phase diagram inferred from energy 
minimisation for a given $U/t_{hop}$ and $n_{exc}$. 
The diagram displays three distinct phases. The boundary 
$n_{exc}^S(U)$ separates the $120^\circ$ phase from the 
spiral phase. The color in the spiral phase indicates 
the ordering wavevector $\vec{Q}$, as shown in the 
inset. The boundary $n_{exc}^{PM}(U)$ marks where the 
magnetic moment size $m$ at the ordering wavevector 
reaches 5\% of its ideal value (0.5), and this 
paramagnetic region (PM) is highlighted in dark grey.
The light grey region also shows a paramagnetic phase 
but with a very small moment size ($<0.025\%$). 
}
\end{figure}
% --------------------------------------------------------

We model the population \(P(x)\) as follows:
\begin{equation}
    P(x) = \begin{cases} 
    a - b x & \text{for } x < 1, \\
    1 - \left(a - b (2 - x) \right) & \text{for } x \geq 1.
    \end{cases}
\end{equation}
This is similar to the analysis in Section IV, but with \(c = 0\) and 
\(a\) fixed at 1, making \(b\) the only free parameter that varies
\(n_{exc}\). 
Populating the 
eigenenergy levels of a cluster with a 
total of $2N_c$ states are well-approximated 
by this simplification, 
which remains valid for \(n_{exc}\) up to 
approximately $30\%$. Within this range, our phase diagram 
encompasses the long-range orders.
We analyze the 
problem on a \(36 \times 36\) lattice for \(U/t_{hop}\) 
ranging from 3 to 20. The resulting phase diagram 
is shown in Fig.9. 

The \(n_{exc} = 0\%\) line 
represents the equilibrium state. It involves the 
following phases: a correlated paramagnetic metal (grey, 
not shown explicitly) up to \(U/t_{hop} \sim 4\), 
followed by an incommensurate magnetic metal (shades of red), 
and finally the \(120^\circ\) insulator (yellow). 
The dash-dotted line indicates the $n_{exc}$ upto
which  
\(120^\circ\) order remains stable. 

For \(U/t_{hop} < 8\), 
the ordering wavevector shifts inward within the Brillouin zone 
(reds). For \(U/t_{hop} > 9\), the ordering wavevector moves 
along the zone boundary (blues). In the range \(U/t_{hop} 
\sim 8 - 9\), the energy landscape exhibits ring-shaped 
minima. As \(n_{exc}\) increases, the energy 
landscape flattens, with the magnetic moment \(m \sim 5\%\) of 
its ideal value (0.5). This is marked by a dotted line in Fig.9, 
where the entire grey zone signifies the paramagnetic phase.

\subsection{Thermalisation}
In mean field theories, often the system does not thermalize.
To address this situation
specific calculations 
have been performed to study thermalisation times 
in Mott insulators. These studies indicate that when a pump pulse 
excites electrons across the gap \(\Delta\), it results in double 
occupancy. The excited electrons then relax through multimagnon emission, 
where each magnon has an energy approximately equal 
to \(J = 4t_{hop}^2/U\). 
Early estimates of the decay time, provided by Strohmaier et al. ~\cite{therm1}, 
suggest it follows the form \(\tau_D \sim \frac{\hbar}{J} 
e^{\left(\alpha \frac{U}{zJ}\right)}\), where \(z\) is the 
lattice coordination number and \(\alpha\) is a coefficient of 
order unity.

The key takeaway from this result is that the time required to emit
multiple `bosons', each with energy \(\sim zJ\), to deexcite an 
electron with energy \(U \gg zJ\), grows exponentially. 
This implies that the emission processes must occur sequentially. 
This conclusion was confirmed through exact diagonalisation 
calculations by Lenarcic and Prelovsek~\cite{therm2}.

In our specific case (\(U/t_{hop} \sim 6.4\)), we have 
\(\Delta/t_{hop} \sim 1\), and given that \(J/t_{hop} \sim 10^{-2}\), 
the decay time is significantly longer than our runtime, 
indicating that thermalisation would be exceedingly slow 
compared to the timescales of our simulations.

\section{Conclusion}

We studied the pump response of the triangular lattice Mott-Hubbard 
insulator in the regime of $120^{\circ}$ order using a combination 
of numerical tools. Spatio-temporal mean-field dynamics (MFD) reveals 
that weak pumping simply reduces the magnitude of  $120^{\circ}$ order.
Upon increasing pump strength, it leads to dynamics where the complete 
suppression of $120^{\circ}$ order is followed by the emergence of 
a spiral state with a smaller moment. At even larger pump strength there is 
destruction of the local moment itself due to saturated double 
occupancy.  Since MFD indicates that a stable upper Hubbard band 
population $n_{exc}$ quickly forms and remains for long time, we used 
this $n_{exc}$ as an input to a variational calculation (VC) to 
confirm the new order that MFD generates, and also construct a larger 
nonequilibrium  $U-n_{exc}$ `phase diagram'. The timescale for the actual emergence of the new order, following the destruction of the $120^{\circ}$ 
state was estimated by using Langevin dynamics (LD) and yields a 
formation time $\sim 10^3-10^4$ times the electronic timescale. Beyond 
our triangular lattice results, this paper demonstrates that a combination 
of MFD, VC, and LD can be used to quickly unveil pump induced emergent 
phases in other gapped systems, for example, charge-ordered or
superconducting systems.
\\

{\it Acknowledgment:} We acknowledge the use of the HPC 
clusters at HRI. SSB and TM were supported in part by an
Infosys award.

\vfill

\newpage

% -------------------------------------------------------------

\end{document}